\documentclass[useAMS,usenatbib]{mn2e}
\usepackage{epsfig}
\usepackage{amsmath}
\begin{document}

\title[The initial mass function of the rich young LMC cluster NGC
1818]{The initial mass function of the rich young cluster NGC 1818 in
the Large Magellanic Cloud}

\author[Q. Liu et al.]{Q. Liu,$^{1,2}$\thanks{E-mail: liuq@bao.ac.cn}
R. de Grijs,$^{3,1}$\thanks{E-mail: R.deGrijs@sheffield.ac.uk}
L. C. Deng,$^{1}$\thanks{E-mail: licai@bao.ac.cn}
Y. Hu,$^{1,2}$ I. Baraffe$^{4}$ and S. F. Beaulieu$^{5}$\\
$^{1}$National Astronomical Observatories, Chinese Academy of
Sciences, Beijing 100012, P. R. China\\
$^{2}$Graduate University of the Chinese Academy of Sciences, Beijing
100049, P. R. China\\
$^{3}$Department of Physics \& Astronomy, The University of Sheffield,
Sheffield S3 7RH\\
$^{4}$CRAL, \'{E}cole Normale Sup\'{e}rieure, 46 all\'{e}e d'Italie,
69007 Lyon, France\\
$^{5}$D\'{e}partement de Physique, de G\'{e}nie Physique et d'Optique
and Centre de Recherche en Astrophysique du Qu\'ebec, Universit\'{e}
Laval, \\ Qu\'ebec, QC G1V 0A6, Canada}

\pagerange{\pageref{firstpage}--\pageref{lastpage}} \pubyear{2008}

\maketitle

\label{firstpage}

\begin{abstract}
We use deep {\sl Hubble Space Telescope} photometry of the rich, young
($\sim$20--45 Myr-old) star cluster NGC 1818 in the Large Magellanic
Cloud to derive its stellar mass function (MF) down to $\sim 0.15$
M$_\odot$. This represents the deepest robust MF thus far obtained for
a stellar system in an extragalactic, low-metallicity ([Fe/H] $\simeq
-0.4$ dex) environment. Combining our results with the published MF
for masses above 1.0 M$_{\odot}$, we obtain a complete present-day
MF. This is a good representation of the cluster's {\it initial} MF
(IMF), particularly at low masses, because our observations are
centred on the cluster's uncrowded half-mass radius. Therefore,
stellar and dynamical evolution of the cluster will not have affected
the low-mass stars significantly. The NGC 1818 IMF is well described
by both a lognormal and a broken power-law distribution with slopes of
$\Gamma=0.46 \pm 0.10$ and $\Gamma \simeq -1.35$ (Salpeter-like) for
masses in the range from 0.15 to 0.8 M$_{\odot}$ and greater than 0.8
M$_{\odot}$, respectively. Within the uncertainties, the NGC 1818 IMF
is fully consistent with both the Kroupa solar-neighbourhood and the
Chabrier lognormal mass distributions.
\end{abstract}


\section{Introduction}

The shape of the stellar initial mass function (IMF) is of great
importance in modern astrophysics. It plays a crucial role in many of
the remaining `big' questions, e.g., the formation and evolution of
the first stars and galaxies. Whether or not the IMF is universal
remains hotly contested (e.g., Scalo 1998, 2005; Eisenhauer 2001;
Gilmore 2001; Kroupa 2007; and references therein). Obtaining the IMF
is challenging because stellar masses cannot be measured directly,
while limitations due to the observational technology used affect
accurate analysis of, particularly, the low-mass stars.

Star clusters, both open and globular clusters, represent ideal
objects to address many astronomical problems because all of their
member stars have the same age and metallicity and are located roughly
at the same distance. Much work has been done on the MFs of globular
clusters (GCs) in the Milky Way. Paresce et al. (2000) found that the
MFs of Galactic GCs are best approximated by a lognormal function,
based on their analysis of the MFs of a dozen Galactic GCs for stellar
masses below 1 M$_{\odot}$. However, since all Galactic GCs are old
($t \ga 10$ Gyr, with typical relaxation times of $\sim 0.1$ Gyr),
they can only provide evolutionary information on a single (long)
time-scale; stellar and dynamical evolution must obviously have
affected the MF at high masses (e.g., de Grijs et al. 2002a,b). The
Large Magellanic Cloud (LMC), on the other hand, is a unique
laboratory for studying star cluster evolution on a range of
time-scales as it contains a large population of rich star clusters
with masses similar to those of Galactic GCs and covering ages from
0.001 to 10 Gyr (e.g., Beaulieu et al. 1999; Elson et al. 1999), thus
making it possible to study clusters at (almost) all evolutionary
stages. Particularly for the rich, young clusters, stellar and
dynamical evolution have not yet affected the MF at low masses, so we
can attempt to obtain the low-mass IMF of these young clusters from
their present-day mass functions (PDMFs). The unprecedented high
spatial resolution of the {\sl Hubble Space Telescope (HST)} allows us
to resolve individual stars in dense star clusters at the distance of
the LMC, $\sim$50 kpc.

\begin{table}
 \centering
 \begin{minipage}{140mm}
\caption{Fundamental parameters of NGC 1818.}
  \begin{tabular}{@{}lcc@{}}
  \hline
  &  & Ref. \\
 \hline
 RA, Dec (J2000)   & $05^{\rm h} 04^{\rm m} 03^{\rm s}$, $-66^\circ 26' 00''$ & 1\\
 $M_V$ (mag)       & $-8.8$                 & 4,6 \\
 log(Age yr$^{-1}$)& $7.65 \pm 0.05$        & 2$^a$ \\
                   & $7.25 \pm 0.40$        & 2$^b$ \\

 [Fe/H] (dex)      & $\sim -0.4$            & 4\\
 $E(B-V)$ (mag)    & 0.03                   & 3 \\
 $(m-M)_{0}$ (mag) & 18.58                  & 3 \\
 Mass (M$_{\odot}$) & $2.8\times 10^{4}$  & 5 \\
 $R_{\rm core}$ (pc)& 2.1 $\pm$ 0.4         & 7 \\
 $R_{\rm hl}$ (pc)  & 2.6 & 8 \\
\hline
\end{tabular}
\flushleft References: 1. de Grijs et al. (2002a); 2. this paper; 3.
Castro et\\ al. (2001); 4. Johnson et al. (2001); 5. Hunter et al.
(1997);\\ 6. van den Bergh (1981); 7. Elson et al. (1989); 8.
Santiago et al.\\ 2001.\\
{\sc Notes:} $^a$ best-fitting main-sequence age (Girardi et al. 2000\\
isochrones), where the age uncertainty originates from the\\
discreteness of the isochrones.\\
$^b$ Average PMS age (Baraffe et al. 1998 isochrones), where the\\
uncertainty represents the most appropriate age range covered\\ by the
PMS stars.
\end{minipage}
\end{table}

NGC 1818 is a young, compact cluster (see Table 1 for the cluster's
fundamental parameters). It has been studied extensively (e.g., Will
et al. 1995; Elson et al. 1998; Johnson et al. 2001; de Grijs et al.
2002a,b,c). Will et al. (1995) studied the cluster's IMF using the
ESO/MPIA 2.2m telescope at La Silla observatory, Chile, but only for
the massive stars ($V\leqslant 22.75$ mag, corresponding to masses
$\ga 1.26$ M$_{\odot}$). de Grijs et al. (2002a,b) obtained the MF
above 1 M$_{\odot}$ using (in part) the same data as analysed in this
paper and concluded that the cluster's MF is largely similar to the
Salpeter (1955) IMF approximation over this mass range. The cluster's
IMF for stellar masses below 1 M$_{\odot}$ is still unknown.

Given the young age of NGC 1818, most of its member stars with masses
below 1.0 M$_{\odot}$ are still on the pre-main sequence
(PMS). Although PMS membership-selection criteria have been
established for the separation of PMS stars from low-mass (zero-age)
main-sequence objects (Park et al. 2000; Sung et al. 2000), using PMS
stars is very challenging on the basis of optical observations alone
because these stars are usually very faint overall and therefore
difficult to detect with most ground-based instruments. The stellar
systems in the LMC are thought to be the best places to search for PMS
stars because they do not suffer from either severe crowding or
significant extinction (Gouliermis et al. 2006a). The evolution of PMS
stars is still rather uncertain (e.g., Baraffe et al. 1998), which
implies that the choice of one's input parameters will significantly
affect the predicted PMS evolution. White et al. (1999) compared six
PMS evolution models and concluded that the models of Baraffe et
al. (1998) resulted in the most consistent ages and masses (Park et
al. 2000). On the basis of the best available evolutionary models
(Baraffe et al. 1998), in this paper we obtain the low-mass IMF (below
1.0 M$_{\odot}$) for the young LMC cluster NGC 1818 based on
high-resolution {\sl HST} data.

Although new evolutionary models for young low-mass stars have
appeared in the literature (see for a review Hillenbrand \& White
2004) since the comparative analysis of White et al. (1999), we
emphasise that the models of Baraffe et al. (1998) use the most
up-to-date equation of state (which was validated on the basis of
high-pressure experiments) and outer-boundary conditions based on
non-grey atmosphere models (see for details Chabrier et al. 2005;
Mathieu et al. 2007). More recent models either use an equation of
state that is more appropriate for solar-type and more massive stars,
but not for the interior conditions of low-mass stars (cf. Yi et al.
2003; Dotter et al. 2008), or approximate grey outer-boundary
conditions which provide incorrect effective temperatures and
luminosities in the presence of molecules (Palla \& Stahler 1999).
The Dotter et al.  (2008) models use the same atmosphere models
(Hauschildt et al.  1999) as Baraffe et al. (1998) while the Siess et
al. (2000) models use similar input physics and are very comparable in
the low-mass stellar domain. Finally, the Chabrier et al. (2000)
models are most suitable for `dusty' conditions, i.e., for objects
with lower masses than discussed here. The Baraffe et al. (1998)
models have been extensively validated observationally for objects of
different ages and masses below 1 M$_\odot$, using different
observational constraints (such as mass-radius and mass-luminosity
relationships, colour-magnitude diagrams, spectra and binary
systems). The only model suite comparable in terms of input physics
and quality are the Siess et al. (2000) models.  However, they do not
include metallicities below $Z=0.01$.

In Section 2 we present the observations and data-reduction steps in
detail. We describe how we obtain the cluster's low-mass IMF in
Section 3. Finally, we discuss our results and provide a summary in
Sections 4 and 5.

\section{Observations and data reduction}

\subsection{Observations}

As part of {\sl HST} programme GO-7307, we have a unique set of
high-quality imaging observations of NGC 1818, obtained with both the
Wide-Field and Planetary Camera-2 (WFPC2) and the Space Telescope
Imaging Spectrograph (STIS). The cluster was part of a carefully
selected LMC cluster sample (see Beaulieu et al. 1999). It has an age
of $\approx 45$ Myr and a mass of $\approx 10^{4}$ M$_{\odot}$ (see
Hunter et al. 1997; Beaulieu et al. 1999; de Grijs et
al. 2002a). WFPC2 is composed of four chips (each containing
800$\times$800 pixels), one Planetary Camera (PC) and three Wide-Field
(WF) arrays. The PC's field of view is about 34$\times$34 arcsec$^{2}$
(with a pixel size of 0.0455 arcsec) and the field of view of each of
the WF chips is about 150$\times$150 arcsec$^{2}$ (with a pixel size
of 0.097 arcsec). The STIS field of view is 28$\times$52 arcsec$^{2}$,
with a pixel size of 0.0507 arcsec.

We obtained WFPC2 exposures through the F555W and F814W filters
(roughly corresponding to the Johnson-Cousins $V$ and $I$ bands,
respectively), with the PC centred on both the cluster core and its
half-mass radius. We have both deep (exposure times of 140 and 300 s
for each individual image in F555W and F814W, respectively) and
shallow (exposure times of 5 and 20 s, respectively) images with the
PC located on the cluster centre (Santiago et al. 2001; de Grijs et
al. 2002a). For the exposures centred on the cluster's half-mass
radius we obtained deep observations with a total exposure time of
2500 s in both filters.

Given that NGC 1818 is observed superimposed on the LMC background
field, it is important to (statistically) subtract the background
stars to obtain clean luminosity and mass functions. We obtained very
deep WFPC2 exposures through the F555W and F814W filters from the {\sl
HST} Data Archive of the general LMC background and of a specific
background region associated with NGC 1818 (see, for more details,
Castro et al. 2001; Santiago et al. 2001; de Grijs et al. 2002a). For
the general background the exposure times were 7800 s (F555W) and 5200
s (F814W) (de Grijs et al. 2002a), while they were 1200 s (F555W) and
800 s (F814W) for the images of the specific background field
associated with the cluster (Santiago et al. 2001; de Grijs et
al. 2002a). Both sets of background fields were significantly deeper
than the targeted cluster observations, hence allowing us to properly
correct for background effects at the faintest magnitudes covered by
our science observations.

We also obtained deep STIS CCD observations in \hbox{ACCUM} imaging
mode through the F28$\times$50LP long-pass filter (central wavelength
$\lambda_{\rm c} = 7230$\AA), with the CCD centred on the half-mass
radius of NGC 1818. The total exposure time of this observation was
2950 s in a set of 5 observations (see also Elson et al.  1999); each
observation was split into two exposures to allow for the removal of
cosmic rays by the data-processing pipeline. The deep STIS data allow
for the construction of very deep luminosity functions; in fact, STIS
(in imaging mode through long-pass filters) is five times more
sensitive for faint red objects than WFPC2 (e.g., de Grijs et
al. 2002a).

\subsection{Data reduction and photometry}

We used the {\sc iraf}/APPHOT\footnote{The Image Reduction and
Analysis Facility ({\sc iraf}) is distributed by the National
Optical Astronomy Observatories, which is operated by the
Association of Universities for Research in Astronomy, Inc., under
cooperative agreement with the US National Science Foundation.}
package to perform aperture photometry. We compared the different
colour-magnitude diagrams (CMDs) for 1-, 2-, 3- and 4-pixel
apertures and found that the 2-pixel aperture was best suited for
stellar aperture photometry in this cluster since it produced the
smallest photometric errors and the tightest main sequence. This
radius corresponds to $\sim$0.09 arcsec for the PC chip, $\sim$0.2
arcsec for the WF chips and $\sim$0.1 arcsec for STIS. It is a
compromise between the need to include the core of the point-spread
function (PSF) but avoid contamination from nearby objects (cf. de
Grijs et al. 2002a).

We emphasise that point-spread function (PSF) fitting and aperture
photometry are both suitable techniques one can use in the type of
environment we are dealing with here. There are pros and cons
associated with either method. Here, we have followed established
practice for NGC 1818 (and our other sample clusters; cf. de Grijs
et al. 2002a,b,c) based on {\sl HST} observations (see Castro et al.
2001; Santiago et al. 2001; Beaulieu et al. 2001; Johnson et al.
2001; de Grijs et al. 2002a,b,c; Hu et al. 2009), for which we
showed that the resulting data quality based on aperture photometry
is robust and sufficient. The key issues are that (i) the sky
background should be as constant as possible (which is met because
of the high-quality {\sl HST} observations available), (ii) we have
carefully determined aperture corrections based on TinyTim simulated
PSF analysis (Krist \& Hook 2001), and (iii) our error bars reflect
the approach used; we base our results on a proper and careful
analysis of the observed signal in terms of the error bars.  The
latter is most crucial and was carefully implemented in our
modeling. The photometric uncertainties of the vast majority of our
individual STIS magnitudes are less than 0.05 to 0.10 mag, while
none of the stars in our final sample have uncertainties greater
than 0.20 mag.

We adopted the relations of Whitmore et al. (1999) to correct the
resulting photometry for the effects of charge-transfer (in)efficiency
(CTE):
\begin{eqnarray}
  Y\mbox{-CTE} &=& 2.3\times10^{-0.256\times\log({\it BKG})}\nonumber\\
    & & \times[1+0.245\times(0.0313-0.0087\log {\it CTS}_{\rm obs})\nonumber\\
    & & \times({\rm MJD}-49471)] ;
\end{eqnarray}
\begin{eqnarray}
  X\mbox{-CTE} &=& 2.5\nonumber\\
    & & \times[1+0.341\times(0.00720-0.0020\log {\it CTS}_{\rm obs})\nonumber\\
    & & \times({\rm MJD}-49471)] .
\end{eqnarray}
\noindent
The total CTE correction is obtained as
\begin{eqnarray}
  {\it CTS}_{\rm cor} &=& (1+\frac{Y\mbox{-CTE}}{100} \times \frac{Y}{800}+\frac{X\mbox{-CTE}}{100} \times \frac{X}{800}) \times \nonumber\\
& & {\it CTS}_{\rm obs},
\end{eqnarray}
where {\it CTS}$_{\rm cor}$ is the number of counts after CTE
correction, {\it CTS}$_{\rm obs}$ is the raw number of counts, $Y$-CTE
is the CTE loss (in per cent) over 800 pixels in the $Y$ direction and
$X$-CTE is the equivalent factor in the $X$ direction, $X$ and $Y$ are
the $x$ and $y$ positions of the star in pixels, {\it BKG} is the mean
number of counts for a blank region of the background field and MJD is
the modified Julian date.

Before applying aperture corrections (ACs), we also used {\sc
iraf/stsdas}\footnote{{\sc stsdas}, the Space Telescope Science Data
Analysis System, contains tasks complementary to the existing {\sc
iraf} tasks. We used version 3.1 for the data reduction performed in
this paper.} to correct for the geometric distortion of the WFPC2
chips. We determined the ACs for our photometry based on the model
PSFs generated by TinyTim (Krist \& Hook 2001). We used single ACs for
the entire chip, because if we change the position where we generate
the TinyTim PSF image the ACs are very similar (positional differences
in the ACs are less than 0.02 mag across the detector). We first
constructed artificial TinyTim PSF images for each passband and each
chip. Next, we measured the flux within a 2-pixel circular aperture
and compared this with the total flux in the artificial PSF, thus
giving us the AC for any given filter/chip combination (listed in
Table 2). We used the same photometric method for the STIS
F28$\times$50LP image.

\begin{table}
 \centering
 \begin{minipage}{140mm}
\caption{Aperture corrections.}
  \begin{tabular}{@{}ccc@{}}
  \hline
 Filter     & Chip & AC (mag) \\
 \hline
 F555W      & PC & 0.424\\
            & WF2 & 0.258\\
            & WF3 & 0.307\\
            & WF4 & 0.274\\
 F814W      & PC & 0.607 \\
            & WF2 & 0.282 \\
            & WF3 & 0.334 \\
            & WF4 & 0.298 \\
 F28$\times$50LP & STIS & 0.506\\
\hline
\end{tabular}
\end{minipage}
\end{table}

We adopted the method of Holtzman et al. (1995) to convert the
aperture-corrected WFPC2 photometry to the standard Johnson-Cousins
$V$ and $I$ passbands:
\begin{eqnarray}
  V &=& -2.5\times\log\dot{C}({\rm F555W})+(-0.052\pm0.007)\times(V-I)\nonumber\\
    & & +(0.027\pm0.002)\times(V-I)^{2}+(21.725\pm0.005)\nonumber\\
    & & +2.5\times\log(GR)
\end{eqnarray}
\noindent and
\begin{eqnarray}
  I &=& -2.5\times\log\dot{C}({\rm F814W})+(-0.062\pm0.009)\times(V-I)\nonumber\\
    & & +(0.025\pm0.002)\times(V-I)^{2}+(20.839\pm0.006)\nonumber\\
    & & +2.5\times\log(GR),
\end{eqnarray}
where $\dot{C}$ is the count rate in 2-pixel apertures and $GR =
1.987, 2.003, 2.006$ and 1.955 for the PC, WF2, WF3 and WF4 chips,
respectively (Holtzman et al. 1995).

Figure 1 shows the CMD of NGC 1818 based on our WFPC2 data. Figure 2
represents the spatial distribution of the stars in the NGC 1818 field
as observed with WFPC2 and STIS (dots and solid rectangular area,
respectively).

\begin{figure}
\includegraphics[width=\columnwidth]{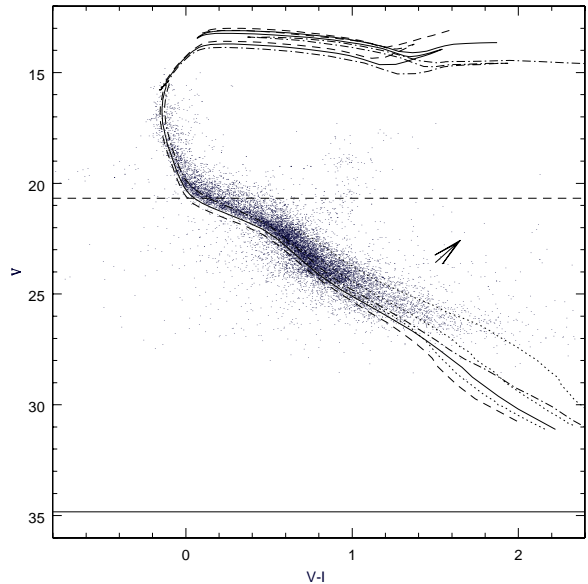}
\caption{{\sl HST}/WFPC2 CMD of NGC 1818. The photometry was
converted to the standard $V$- and $I$-band filters using Eqs. (4)
and (5). From bottom to top (in the direction of the arrow) the
three dotted lines represent isochrones (Baraffe et al. 1998) of
log(Age yr$^{-1}$) = 7.25 (best fit to the average PMS) and
metallicities $Z= 0.0019, 0.006$ and 0.019, respectively. The
dashed, solid and dot-dashed lines represent Padova isochrones
(Girardi et al. 2000) of log(Age yr$^{-1}$) = 7.65 (best fit to the
main sequence) and metallicities $Z= 0.004, 0.008$ and 0.019,
respectively. The horizontal dashed and solid lines represent,
respectively, the upper and lower magnitude limits to the
colour-magnitude space covered by our STIS observations.}
\label{CMD(model)}
\end{figure}

\begin{figure}
\includegraphics[width=\columnwidth]{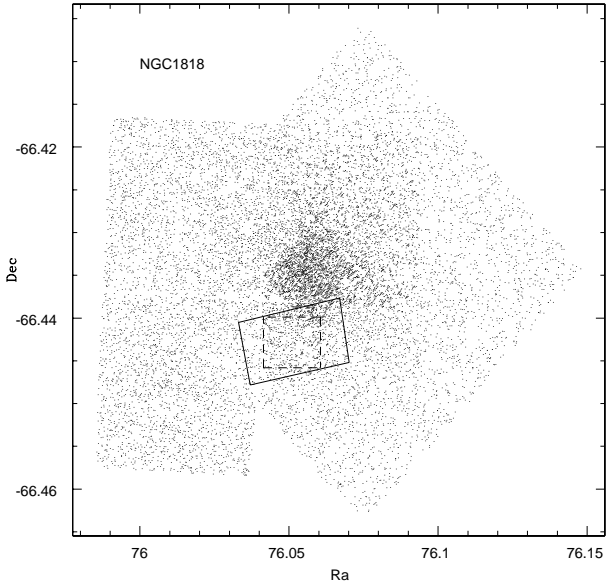}
\caption{Spatial distribution of the stars in NGC 1818. The dots
represent the stars detected in the WFPC2 observations. The solid
rectangular area indicates the area covered by our STIS observations
and the short-dashed square delineates the area used for further
analysis in this paper (see Section 3.3). \label{spatial
diagram(WFPC2)}}
\end{figure}

\subsection{Completeness}

Because of the cluster's stellar density gradient, one of the most
difficult problems for MF derivation involves completeness correction,
which is normally a function of position within a cluster. We used a
similar method of completeness correction as de Grijs et al. (2002a).
They computed the corrections in circular annuli around the cluster
centre. However, we simply computed the equivalent corrections for the
entire chip for the exposures centred on the cluster's half-mass
radius, as well as for the entire STIS chip, because both sets of
observations are centred on roughly the same position and the area
around the half-mass radius is not as crowded as the cluster
centre. The effects of sampling incompleteness are constant across our
STIS field, within the observational uncertainties (see also de Grijs
et al. 2002a). The same method was used for the completeness
corrections applied to the background field, although for the
magnitude range of interest the completeness of the background field
is close to unity.

\begin{figure}
\includegraphics[width=\columnwidth]{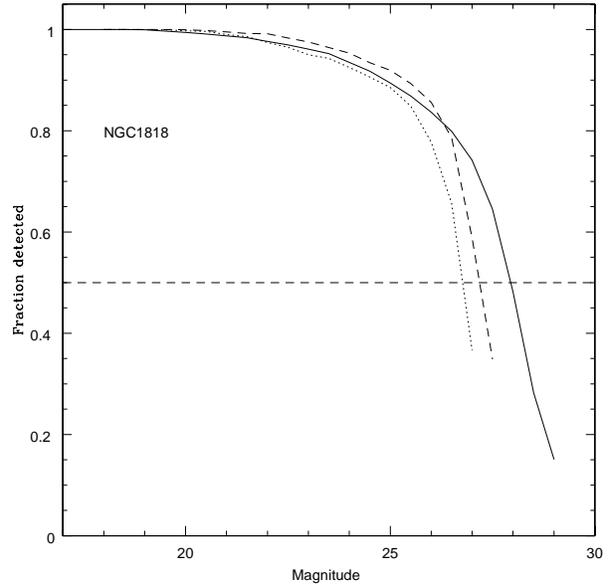}
\includegraphics[width=\columnwidth]{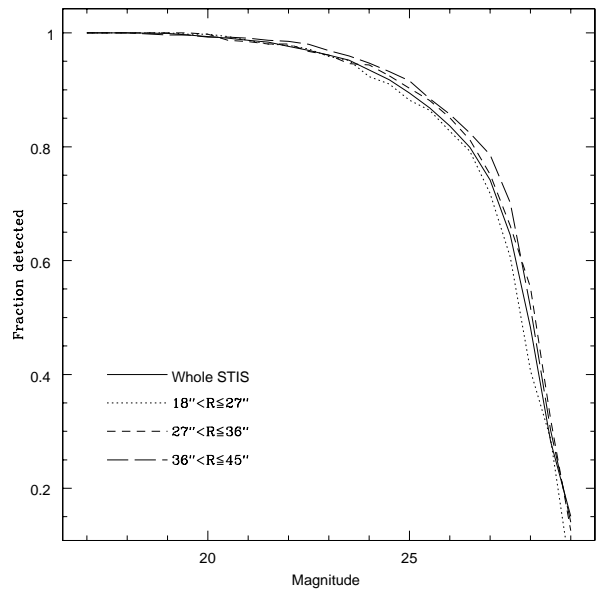}
\caption{Completeness curves for NGC 1818. (Top) Solid line: STIS;
dotted line: PC (F555W); short-dashed line: PC (F814W). The horizonal
dashed line represents the 50 per cent completeness limit. (Bottom)
Radially-dependent completeness curves of the STIS observations.
\label{completecurve}}
\end{figure}

We added an area-dependent number of artificial sources of Gaussian
shape to each chip for input magnitudes between 15.0 and 30.0 mag, in
steps of 0.5 mag. We then applied the same photometric method to the
fields including both the cluster stars and the artificial sources, to
quantitatively assess how many artificial stars we can detect after
correction for chance blends and superpositions. Figure 3 shows the
completeness curves of the STIS and PC chips. The bottom panel shows
that the STIS completeness curves are not significantly radially
dependent. We therefore used the completeness corrections for the STIS
data as a whole (the variation of the different curves near the 50 per
cent completeness limit is negligible in relation to the observational
uncertainties). The results shown have all been corrected for blending
of multiple randomly placed artificial stars and for superposition of
artificial and real stars in the cluster region. For our analysis in
the remainder of this paper we only consider magnitude ranges that are
$\ge50$ per cent complete.

\section{Analysis}

\subsection{Evolutionary model, age and metallicity}

Since the observations were obtained in two WFPC2 and one STIS
filter, we obtain three LFs, in the F555W, F814W and F28$\times$50LP
bands. These must be converted to a common parameter (e.g., mass)
for comparison purposes. de Grijs et al. (2002b) studied the MF of
NGC 1818, based on the WFPC2 data, and adopted three mass-luminosity
(ML) conversions for stellar masses above $\sim$1.0 M$_{\odot}$, for
solar and subsolar metallicities and a cluster age of 25 Myr. We can
reach much lower masses using the STIS data because of the much
longer exposure times employed. All of the stars detected by STIS
are faint and (hence) of low mass (assuming that they are cluster
members), as is apparent from a direct comparison of the stars in
common on the shallower WFPC2 exposures and the deep STIS frames
centred on the cluster's half-mass radius. For the young age of the
cluster this implies that most of these objects are likely PMS
stars. For the stars in common with the WFPC2 observations (the
brighter stars detected on the STIS frames), this is supported by
their loci in the CMD, because many stars are located above the
zero-age main sequence (ZAMS) of NGC 1818, at least for low masses
and on the basis of WFPC2 photometry alone (see Fig. 1). This
distribution of the photometric data points is not due to biases
caused by photometric uncertainties.

White et al. (1999) concluded that the models of Baraffe et al.
(1998) result in the most consistent age and mass estimates, on the
basis of a detailed comparison of six PMS evolutionary models. We
therefore adopted the Baraffe et al. (1998) models for low-mass (PMS
and ZAMS) stars and recalculated the models for the specific filters
used in this paper and for a more extended range of metallicities and
higher stellar masses (up to 1.4 M$_{\odot}$). Although a comparison
of the Baraffe et al. (1998) PMS and the Girardi et al. (2000) main
sequence models with our observational CMD appears to imply that up to
88 per cent of the stars detected by STIS ($\sim \in$ [22.6,27.8] mag)
may be PMS stars, it is difficult to ascertain whether or not a given
star is on the PMS, so the full, extended mass coverage of the Baraffe
et al. (1998) models significantly aids in the interpretation of our
results. In addition, we added photometry in the F28$\times$50LP
filter to the model calculations.

\begin{figure}
\includegraphics[width=\columnwidth]{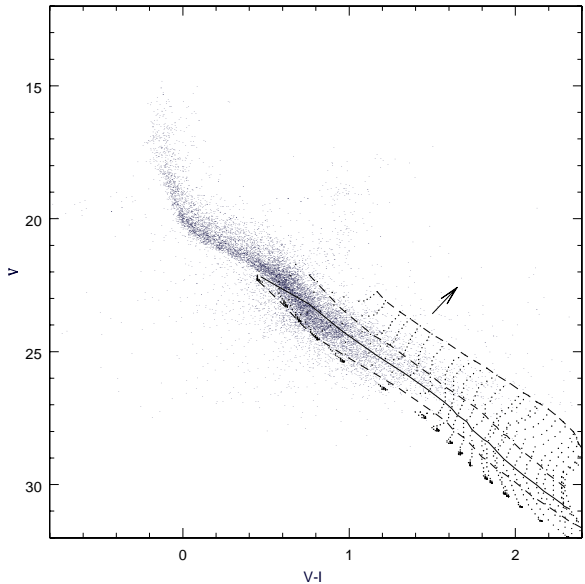}
\caption{Baraffe et al. (1998) models with $Z$= 0.006 overlaid on
the cluster's CMD. The dotted lines are the PMS evolutionary tracks.
The three dashed lines represent isochrones of log(Age yr$^{-1}$) =
7.65, 6.85 and 6.0, respectively (in the direction of the arrow).
The solid line represents an isochrone of log(Age~yr$^{-1}$) = 7.25,
which traces the most appropriate average age for the PMS stars in
the cluster. } \label{PMS.CMD}
\end{figure}

Although we cannot use CMD analysis to attack the PMS uncertainties
for the bulk of the STIS data, we know that the area observed by
STIS overlaps with the WFPC2 observations centred on the cluster's
half-mass radius. Many (more than 1000 after completeness
correction) of the faint stars in the STIS image are also detected
in our WFPC2 observations and most of the stars in common between
the two catalogues are PMS stars. As clearly shown in Fig.  4, the
PMS stars observed by WFPC2 seem to exhibit a scatter in age (e.g.,
compare the youngest and oldest PMS isochrones in Fig. 4 with the
data points).

To derive the age of PMS stars, it is common practice to compare their
loci on the CMD with respect to the ZAMS (e.g., Park et al. 2000;
Gouliermis et al. 2006a). This can be done for the brighter stars in
the STIS field of view using the WFPC2 observations in common. Using
the CMD of the low-mass stars observed with WFPC2 (Fig. 4), we
realized that the PMS stars are distributed in a rather narrow region
(see the detailed discussion below) and that most stars are located
parallel to the main sequence (Park et al. 2000). Therefore, we can
assume that all low-mass PMS stars detected by STIS have a mean age
(and spread) centred on the `PMS sequence', so that a MF for the PMS
stars can be derived. Older and younger age sequences spanning the
full CMD loci of the PMS stars detected in the WFPC2 data are used to
measure the uncertainties associated with using this method. This way,
the MF derived from the low-mass end of the WFPC2 data can be extended
to the much deeper STIS photometry, by adopting a well-justified mean
age (and uncertainty) and thus a specific ML conversion. This will
allow us to probe well into the subsolar-mass regime at a hitherto
unexplored (subsolar) metallicity.

We adopted the Padova isochrones (Girardi et al. 2000) to fit the
main-sequence ridge line in the CMD and the Baraffe et al. (1998)
evolutionary models to fit the PMS loci. In the direction of the arrow
in Fig. 1, the dashed, solid and dot-dashed lines represent the best
Padova isochrone fits for metallicities, $Z=0.004, 0.008$ and 0.019
(Z$_\odot$), for log(Age yr$^{-1}$) = 7.65 ($\simeq 45$ Myr). The
$Z=0.008$ isochrone provides the best fit to the main sequence of NGC
1818. However, the Baraffe et al. (1998) model suite does not include
this metallicity, so we compare their models of $Z=0.0019, 0.006$ and
0.019 (dotted lines in Fig. 1, from bottom to top in the direction
indicated by the arrow). The difference between the models with
$Z=0.006$ (Baraffe et al. 1998) and 0.008 (Girardi et al. 2000) is
negligible, so we will use the Baraffe et al. (1998) $Z=0.006$ model
for our PMS analysis. The dotted lines in Fig. 4 are the corresponding
evolutionary tracks for low-mass stars with $Z=0.006$.  In the
direction of the arrow in Fig 4, the dashed lines represent isochrones
of $Z=0.006$ and log(Age yr$^{-1}$) = 7.65, 6.85 and 6.0.  Any stars
located above the middle dashed line are most likely affected by
significant photometric scatter (note that the effects of binary stars
are small compared to the observed width of the CMD; cf. Hu et
al. 2009). We therefore determine the mass of the low-mass stars based
on the adopted model of log(Age yr$^{-1}$) = 7.25 (solid line in
Fig. 4) as it represents the average age of the best-fitting
isochrones.

\subsection{Background subtraction}

As already discussed by Castro et al. (2001), an old red-giant
population and an intermediate-age red-clump population are clearly
seen in the CMD of NGC 1818. These older components can only be
interpreted as field stars in the LMC's disc. To get a clean MF, the
field-star contamination must be subtracted. There are two relevant
sets of background observations obtained with WFPC2 available in the
{\sl HST} Data Archive, a background field associated with the cluster
itself and a general LMC field. We decided to use the general field
because it provides deeper photometry than the background associated
with the cluster. (The 50 per cent completeness limit for the general
LMC field was determined at $m_{\rm F555W} \simeq 27.5$ mag, while
the equivalent limit for the field associated with the cluster occurs
at $m_{\rm F555W} \simeq 26.3$ mag.)

Castro et al. (2001) adopted isochrones of old age to fit the
background population (see their fig. 5), which implies that the NGC
1818 main sequence is severely contaminated by the low-mass
main-sequence field stars. It is therefore of the utmost importance to
carefully subtract this contaminating population. In addition, the
stars detected in the general background field are all of low
mass. Because of the very long exposure time of the general background
field (7800 and 5200 s in F555W and F814W, respectively), all
high-mass stars are saturated. The mass distribution in the field is
therefore obviously different from the cluster MF (cf. the solid curve
in Fig. 5) and must be subtracted carefully (and in a statistical
sense) from the cluster MFs.

\begin{figure}
\includegraphics[width=\columnwidth]{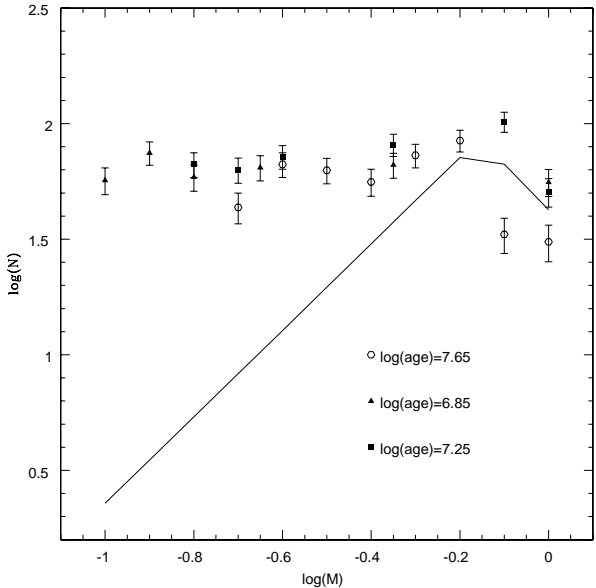}
\caption{Mass distribution adopting Baraffe et al. (1998) ML relations
for different ages in the common WFPC2/STIS area (after background
subtraction); the solid curve shows the mass distribution of the
general background.}
\label{MF.7.65.vs.6.85}
\end{figure}

We do not have access to a background field in the STIS
F28$\times$50LP passband. Instead, we use the F814W photometry as a
proxy for the general background in the STIS field of view because its
effective wavelength is closest to the central wavelength of the
F28$\times$50LP band. Even so, the observations of the general
background are still not as deep as the STIS data. Gouliermis et
al. (2006b) suggested that the stellar mass distribution in the field
of the LMC follows a broken power-law distribution, based on their
study of the general background field of LMC's inner disc. In
addition, the stellar mass function below 1.0 M$_{\odot}$ is generally
considered to be well approximated by both a broken power-law (e.g.,
Kroupa 2001; Kroupa et al. 2003; Covey et al. 2008) and a lognormal
distribution (e.g., Paresce et al. 2000; Chabrier 2003; Andersen et
al. 2008; Hennebelle \& Chabrier 2008). Therefore, we adopted both a
power-law (with a slope $\Gamma$=1.87$\pm$0.06, where the IMF,
$\xi$(m) $\propto$ $m^{\Gamma}$) and a lognormal function to
approximate and extrapolate the general-background MF for masses
between 0.15 and 0.63 M$_{\odot}$, the mass range where we only have
STIS photometry.

\subsection{The mass function}

For WFPC2 we focus on the observations centred on the cluster's
half-mass radius, because they are much deeper (hence reaching lower
masses) than the pointings centred on the cluster core, and well
outside the most crowded region. As WFPC2 covers a much larger field
of view than STIS, we must carefully choose the area in common for a
proper comparison of the different MFs. We used the dashed square
region shown in Fig. 2 for this purpose, for which we calculated the
low-mass MFs in NGC 1818 based on both the high-resolution PC data and
the STIS field of view. The MF from the common area on STIS is
identical to that from the full STIS field of view and thus the mass
distribution of the common area is fully representative of the
cluster's MF as a whole at this distance from the cluster centre.

\begin{figure}
\includegraphics[width=\columnwidth]{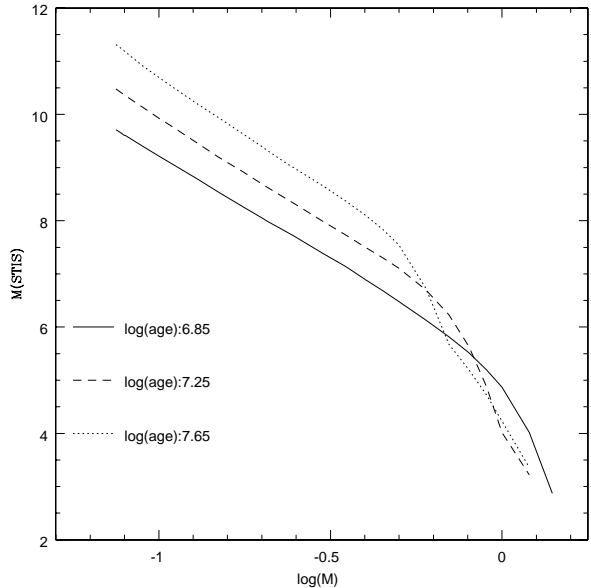}
\caption{Mass-luminosity relations for different ages based on the
Baraffe et al. (1998) models.} \label{ML.relation}
\end{figure}

We assume that all low-mass stars observed with STIS have the same age
(see Sect. 3.1). The most likely age range for the entire stellar
population runs from log(Age yr$^{-1}$) = 6.85 to 7.65. The models for
different ages are characterized by different ML relations (see
Fig. 6). Adopting different models for our luminosity-to-mass
conversion will therefore lead to variations in the mass distribution,
including in the mass range and the number of stars in each mass
interval (see Fig. 5).

\begin{figure}
\includegraphics[width=\columnwidth]{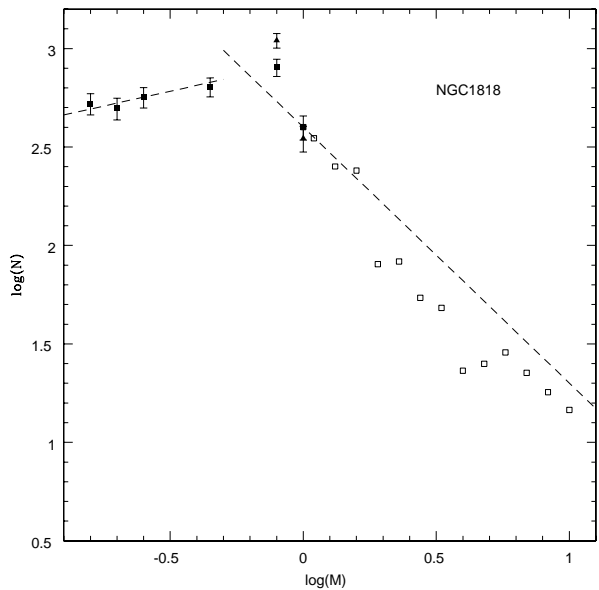}
\caption{Complete MF of NGC 1818, based on background subtraction
assuming a power-law IMF at low masses. Open squares: high-mass MF (de
Grijs et al. 2002b); solid squares: STIS MF (this paper); triangles:
WFPC2 MF (this paper). The dashed lines represent the standard Kroupa
(2001) IMF.} \label{IMF}
\end{figure}

\begin{figure}
\includegraphics[width=\columnwidth]{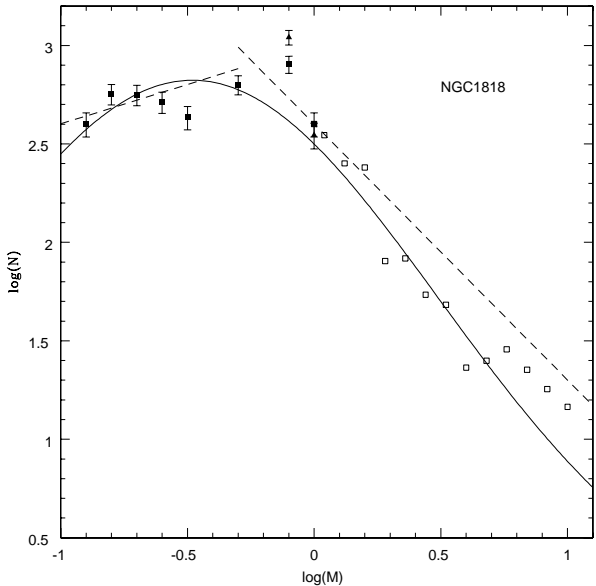}
\caption{Complete MF of NGC 1818, based on background subtraction
assuming a lognormal IMF at low masses. The symbol coding is as in
Fig. \ref{IMF}. The dashed lines show the standard Kroupa (2001) IMF
while the solid line represents the best-fitting lognormal
distribution.} \label{IMF2}
\end{figure}

The WFPC2 F814W filter has a central wavelength ($\lambda_{\rm c} =
8012$\AA) which is much closer to that of the STIS F28$\times$50LP
passband ($\lambda_{\rm c} = 7230${\AA}) than the WFPC2 F555W filter
($\lambda_{\rm c} = 5439${\AA}). The number of stars detected in the
F814W observations is closer to that found in the F28$\times$50LP
frames than the stellar numbers in the F555W observations. Figures 7
and 8 show the mass distributions obtained on the basis of the
observations in the two filters (F814W and F28$\times$50LP for the
WFPC2 and STIS data, respectively) in the common area. The difference
between both figures relates to the assumption adopted regarding the
shape of the low-mass IMF for background subtraction: in Fig. 7 we
assumed a power-law IMF, while in Fig. 8 a lognormal distribution was
imposed.\footnote{The lognormal mass function for stellar masses below
$m_\ast = 1$ M$_\odot$ superimposed on the data in Fig. 8 is given by
\begin{equation}
f(\log m_\ast)= (2.82\pm0.12) \exp \Bigl[-\frac{(\log m_\ast-\log
    0.33^{+0.12}_{-0.09})^2}{2 \times (1.03\pm0.14)^2} \Bigr].
\end{equation}}
The solid squares are based on the STIS data and the triangles
originate from the PC. For masses below 0.8 M$_{\odot}$, the STIS
observations yield more low-mass stars than WFPC2. This may reflect
the true mass distribution of the low-mass stars in NGC 1818. For
stellar masses greater than 0.8 M$_{\odot}$ we use the results from
the WFPC2 observations, since the STIS images are saturated at these
bright magnitudes. At $\log({\rm Mass}/{\rm M}_\odot)=-0.1$ and 0.0,
where stars were detected on both the WFPC2 and STIS chips, we found
1153 stars in common between the two instruments (see Figs. 7 and
8). All WFPC2 stars used to derive the mass function at high masses
are main-sequence stars, while of the STIS-detected stars below 0.8
M$_{\odot}$, $\sim 2226$ are PMS candidates (see for details Table
3). We combined the mass distribution from the two sets of
observations to construct a clean MF spanning the full range from 0.15
to 1.25 M$_{\odot}$, as shown in Figs. 7 and 8.

\section{Discussion}

de Grijs et al. (2002b) studied the PDMF of NGC 1818 for masses above
$\sim$1 M$_{\odot}$ based on three ML relations. The results from the
different ML relations appear similar, with a PDMF slope closely
approximated by the Salpeter (1955) IMF slope ($\Gamma = -1.35$) for
the cluster as a whole. In this paper, we cover a mass interval in
common with their results, so that we can construct a complete,
combined MF for NGC 1818. We adopt the results from de Grijs et
al. (2002b) based on the Kroupa, Tout \& Gilmore (1993) ML relation
(because this IMF extends to the highest stellar masses in NGC 1818)
and combine their MF with our own determination of the PDMF for lower
masses to obtain a complete MF for stellar masses above 0.15
M$_{\odot}$ (Figs. 7 and 8).

The usual method to derive the ages and masses of PMS stars is based
on their loci in the CMD (Gouliermis et al. 2006a; Park et al.
2000). Although this approach is not possible for our STIS
observations, we can use {\it a priori} information to help us
constrain the most likely age range. Park et al. (2000) analysed the
PMS stars in the young cluster NGC 2264 and found that most were
distributed along a sequence parallel to the main sequence. We can use
this information to justify our assumption that all PMS stars detected
by STIS have the same age, which we obtain by fitting the Baraffe et
al. (1998) evolutionary models to the observed sequence.

\begin{figure}
\includegraphics[width=\columnwidth]{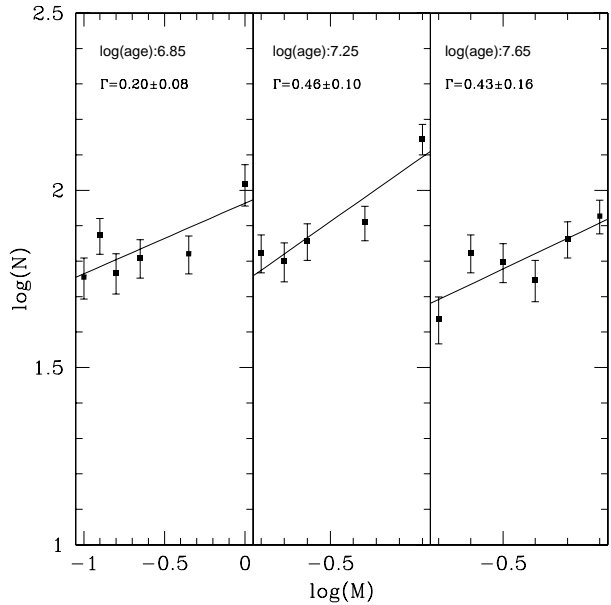}
\caption{NGC 1818 MF based on ML relations for different ages.}\label{IMF3}
\end{figure}

We already concluded that -- based on the WFPC2 observations -- the
most likely age range spanned by the PMS stars in NGC 1818 covers the
interval from log(Age yr$^{-1}$) = 6.85 to 7.65. This, combined with
the symmetrical spread of the data points between these age
boundaries, leads us to adopt a mean age for the NGC 1818 PMS stars of
log(Age yr$^{-1}$) = 7.25. The resulting MF for the youngest age
constraint, combining the STIS and PC data, is shown in the left-hand
panel of Fig. 9 and tabulated in Table 3; the right-hand panel shows
the equivalent MF for the oldest age constraint. For the youngest age
constraint, the best fit to the power-law slope of the MF gives
$\Gamma = 0.20 \pm 0.08$.  On the other hand, for log(Age yr$^{-1}$) =
7.65, $\Gamma = 0.43 \pm 0.16$.

\begin{table}
\centering
\begin{minipage}{140mm}
\caption{NGC 1818 mass function.}
\begin{tabular}{@{}cccc@{}}
\hline
$\log(m_\ast/{\rm M}_\odot$) & $N$ & $\log(m_\ast/{\rm M}_\odot$) & $N$ \\
\hline
$-$0.80& 525 & 0.36 & 83\\
$-$0.70& 497 & 0.44 & 54\\
$-$0.60& 566 & 0.52 & 48\\
$-$0.35& 638 & 0.60 & 23\\
$-$0.10&1099 & 0.68 & 25\\
  0.00 & 399 & 0.76 & 29\\
  0.04 & 350 & 0.84 & 23\\
  0.12 & 252 & 0.92 & 18\\
  0.20 & 240 & 1.00 & 15\\
  0.28 &  80 \\
\hline
\end{tabular}
\end{minipage}
\end{table}

Our results rely on the assumption that the cluster's mass function
consists of single stars. This seems a reasonable assumption because
the effect of unresolved binarity is expected to be very small.
Kerber \& Santiago (2006) analysed the effect of binaries on MFs
containing different binary fractions and characterised by different
MF slopes with $\alpha$ down to 0.8. Our slope corresponds to
$\alpha$=0.54. The binary fraction at these low masses is poorly
known. Marchal et al.(2003) and Delfosse et al. (2004) obtained a
binary fraction of about 30 per cent for this low-mass range. L.
Kerber (priv. comm.) kindly recomputed the tests done in Kerber \&
Santiago (2006) and added a binary fraction of 30 per cent, with
$\alpha$ down to 0.4. The resulting MF (which is, strictly speaking,
a system MF and not a single-star MF) characterised by a binary
fraction of 30 per cent is identical to one with no binarity over
the entire stellar mass range of interest, {\it within the
uncertainties associated with our results,} so that we conclude that
binarity does not play a crucial role in our results, for neither
random pairing nor constant mass-ratio binaries. (Clearly, binaries
will affect the final mass function, but the effect is expected to
be small or negligible in relation to the uncertainties. Hu et al.
[2009] derived $\sim$60 per cent binarity for NGC 1818 for the more
massive F-type stars.)

We used {\sc iraf}/APPHOT for our photometry, leaving the `sharpness'
and `roundness' parameters -- used to weed out non-astronomical
objects -- to the default values, as this resulted in a `clean' sample
of stars in our field of view. Hu et al.  (2009) appropriately adopted
background fields observed with WFPC2 to subtract the background of
their WFPC2 data. Although we could not obtain background CMDs of
similar depth as our STIS observations, the WFPC2 and STIS fields
cover a common area on the CMD (see Fig. 1). We carefully compared the
loci in common of the CMDs obtained from the various observations
(i.e., science data and background fields) and found that there were
few stars beyond the main-sequence ridge line of the background
field. We therefore adopted an appropriate method to extend the MF of
background to a similar depth as the STIS observations (see details in
Section 3.2), i.e., our background subtraction was done as properly as
possible with the data at hand and hence any remaining Malmquist bias
is expected to be minimal.

The stellar mass-dependent dynamical properties of NGC 1818 were
discussed by de Grijs et al. (2002b). The half-mass relaxation
time-scale of cluster stars with masses below 1.0 M$_{\odot}$ is of
order a few times $10^8$ to $10^9$ yr, i.e., much greater than the age
of the cluster. Although dynamical evolution in the cluster core may
be up to 10 to 20 times faster (cf. de Grijs et al. 2002b), at the
half-mass radius the STIS data used here probe a relatively
quiescent,\footnote{Note that although intuitively one would expect
the lower-mass stars to be ejected from the cluster on short
time-scales, the violent encounters between stars leading to ejection
occur predominantly in cluster cores. The environmental conditions at
a cluster's half-mass radius are quiescent, however, particularly for
a relatively extended (for its age) cluster like NGC 1818 (cf. de
Grijs et al. 2002c, their fig. 1).} representative sample of stellar
masses that will have been affected negligibly by dynamical
evolution. Note that the loss of some small fraction of low-mass
stars in unavoidable, even on the relatively short time-scale probed
here. However, given the relatively small mass range probed by our
STIS observations, it is unclear to what extent (if any) this is
important nor whether to expect a differential effect as a function of
stellar mass (e.g., Kroupa 2008, his section 3.2).

This is the first time, to our knowledge, that anyone has probed the
stellar MF to such low masses in an extragalactic, low-metallicity
environment. This makes it difficult to compare our results with other
relevant publications. However, we note that Paresce et al. (2000)
found that the IMF of Galactic GCs is best fit by a lognormal
function, a result based on analysis of the IMF below 1 M$_{\odot}$ of
a dozen GCs in the Milky Way. This conclusion is also consistent with
Chabrier (2003). In an extragalactic environment, Chiosi et al. (2007)
studied the IMF of three young clusters in the Small Magellanic Cloud
(SMC), but their observations did not allow them to probe down to
similarly low masses as done in this paper; they only reach stellar
masses down to $\sim$0.7 M$_{\odot}$. However, Da Rio et al. (2009)
recently probed the stellar mass function of the stellar association
LH 95 in the LMC down to 0.43 M$_\odot$ and found a similar broken
power-law MF based on their {\sl HST} data.

We emphasize that the MFs based on the youngest and oldest age
constraints do not reflect the true underlying mass distribution, but
they provide a robust handle on the uncertainties associated with our
MF analysis (including the expected effects of the cluster's binary
stellar population; cf. Hu et al. 2009).  Instead, the MF for log(Age
yr$^{-1}$) = 7.25 (middle panel of Fig.  9) is much closer to true
mass distribution. It follows a power-law distribution with a slope of
$\Gamma = 0.46 \pm 0.10$ for masses $\leq 0.8$ M$_{\odot}$; the MF
follows the Salpeter mass distribution for masses $> 0.8$ M$_{\odot}$
(Figs. 7 and 8).\footnote{For comparison, for the low-mass MF of
Fig. 8, where we subtracted the background field stars assuming a
lognormal distribution in mass, the equivalent slope is $\Gamma = 0.40
\pm 0.14$, which is identical within the uncertainties.}

Although general agreement exists regarding a `universal'
Salpeter-like stellar IMF for masses greater than $\sim$1.0
M$_{\odot}$ in any environment that is sufficiently well populated
(e.g., Kroupa 2001, 2007; Chabrier 2003; Chiosi et al. 2007), few
studies have managed to constrain the IMF well below 1 M$_{\odot}$
(cf. Paresce et al. 2000; Kroupa 2001, 2007; Chiosi et al. 2007; Covey
et al. 2008). Most current low-mass IMF studies find broken power-law
or lognormal mass distributions; the exact functional form of the
low-mass IMF is still a matter of debate, however. Kroupa (2001)
studied the Galactic-field IMF down to 0.01 M$_{\odot}$ and found a
three-part power-law distribution with turnovers at $\sim$0.08 and
$\sim 0.5$ M$_\odot$ (see also Covey et al. 2008).  Chiosi et
al. (2007) obtain a similar mass function from their analysis of three
clusters in the SMC. Andersen et al. (2008) combine the low-mass
stellar mass distributions for seven star-forming regions and conclude
that the composite IMF is consistent with a lognomal mass function.

Our MF slope for NGC 1818 above 1.0 M$_{\odot}$ (see Figs. 7 and 8 and
details in de Grijs et al. 2002b) is consistent with a `universal'
Kroupa IMF. Within the uncertainties, the MF slope we find for lower
masses ($\la$1.0 M$_\odot$) is similar to that of Kroupa (2001),
supporting a possible `universal' low-mass IMF as well. Given the
intrinsic fluctuations in NGC 1818 and the associated uncertainties,
its turn-over mass is fully consistent with the equivalent mass of 0.5
M$_\odot$ suggested by Kroupa (2001). The resolution in mass of our
data is insufficient to speculate on the cause of any differences. We
re-emphasize here that a satisfactory fit to the low-mass IMF of NGC
1818 can also be obtained on the basis of a lognormal mass function,
although significantly broader than the Chabrier (2003) IMF.

Application of different ML relations will result in different
masses. From Fig. 6 we see that for masses greater than 0.63
M$_{\odot}$ the three ML relations for NGC 1818 are similar, implying
that the shape of the MFs for different age assumptions is similar
(see Fig. 5). For masses below 0.63 M$_{\odot}$ the three ML relations
are parallel. The shape derived from models of different ages should
therefore be the same within the uncertainties, although the number of
stars in each mass interval is different. From Figs. 5 and 9 we find
that the shapes of the MFs for different ages are similar, but the
slopes vary to some extent.

Although we derived the NGC 1818 MF from 0.15 to 1.25 M$_{\odot}$ in
this paper, this is not necessarily the {\it initial} MF as
evolutionary effects, such as (dynamical) mass segregation (de Grijs
et al. 2002a,b,c), may have modified the cluster's IMF. However, given
that NGC 1818 is very young ($\sim$45 Myr), stellar and dynamical
evolution will not have affected the IMF significantly at the lowest
masses. In addition, the location of the common WFPC2/STIS area is far
outside the crowded centre of NGC 1818 so that dynamical evolution is
unlikely to have modified the IMF to any significant extent at these
radii and we can therefore confidently consider the observed MF of the
low-mass stars in NGC 1818 as its IMF.

\section{Summary and conclusions}

In this paper we use deep {\sl HST} WFPC2 and STIS photometry of the
rich young cluster NGC 1818 in the LMC to derive the stellar MF for
low-mass stars down to $\sim 0.15$ M$_\odot$. To the best of our
knowledge, this is the deepest MF thus far obtained for a stellar
system in an extragalactic, low-metallicity environment (although
somewhat -- but not quite -- rivalled by the recent work of Da Rio et
al. 2009).

Combining our results with the MF for stellar masses above 1.0
M$_{\odot}$ derived by de Grijs et al. (2002b), we obtain a complete
PDMF of NGC 1818. This PDMF is most likely a good representation of
the cluster's IMF, particularly at low masses. This is so because NGC
1818 is very young and the observations are centred on a field at the
cluster's uncrowded half-mass radius, so that stellar and dynamical
evolution of the cluster is unlikely to have significantly affected
the NGC 1818 IMF at low masses. Adopting a Kroupa-type power-law mass
distribution as a convenient fitting tool, the IMF in NGC 1818 is best
described by a broken power-law distribution with slopes of
$\Gamma=0.46 \pm 0.10$ and $\Gamma \simeq -1.35$ (Salpeter-like) for
masses in the range from 0.15 to 0.8 M$_{\odot}$ and greater than 0.8
M$_{\odot}$, respectively. Our derived IMF is therefore fully
consistent, within the uncertainties, with the `standard' Kroupa
(2001) broken power-law IMF for the solar neighbourhood. Given the
observational uncertainties, the low-mass IMF is also well
approximated by a lognormal distribution. At the present time, we
cannot robustly distinguish between either functional form. To do so,
one would need to probe down to the stellar/brown-dwarf transition
region. This is, however, very challenging to achieve for clusters at
Magellanic Cloud distances.

\section*{Acknowledgments}

This work was supported by the National Natural Science Foundation of
China under grant No. 10573022 and by the Ministry of Science and
Technology of China under grant No. 2007CB815406. RdG acknowledges
partial financial support from the Royal Society in the form of a
UK--China International Joint Project. We thank Leandro Kerber for his
insights in and simulations of the effects of varying binarity on our
results. This paper is based on archival observations with the
NASA/ESA {\sl Hubble Space Telescope}, obtained at the Space Telescope
Science Institute, which is operated by the Association of
Universities for Research in Astronomy, Inc., under NASA contract NAS
5-26555. This research has made use of NASA's Astrophysics Data System
Abstract Service.

\label{lastpage}


\begin{thebibliography}{}
\bibitem[]{} Andersen M., Meyer M. R., Greissl J., Aversa A., 2008,
        ApJ, 683, L183
\bibitem[\protect\citeauthoryear{Baraffe} {1998}]{b1}
        Baraffe I., Chabrier G., Allard F., Hauschildt P. H., 1998, A\&A,
        337, 403
\bibitem[\protect\citeauthoryear{Beaulieu} {1999}]{b1} Beaulieu S. F,
        Elson R. A. W., Gilmore G. F., Johnson R. A., Tanvir N.,
        Santiago B. X., 1999, in: New Views of the Magellanic Clouds,
        Chu Y.-H., Suntzeff N., Hesser J., Bohlender D., eds., IAU
        Symp. 190, (ASP: San Francisco), p. 460
\bibitem[\protect\citeauthoryear{Castro} {2001}]{b2}
        Castro R., Santiago B. X., Gilmore G. F., Beaulieu S.,
        Johnson R. A., 2001, MNRAS, 326, 333
\bibitem[\protect\citeauthoryear{Chabrier} {2003}]{b3}
        Chabrier G., 2003, PASP, 115, 763
\bibitem[\protect\citeauthoryear{Chabrier et
al.}{2000}]{2000ApJ...542..464C} Chabrier G., Baraffe I., Allard F.,
Hauschildt P., 2000, ApJ, 542, 464
\bibitem[\protect\citeauthoryear{Chabrier et al.}{2005}]{b3}
Chabrier G., Baraffe I., Allard F., Hauschildt P. H., 2005, in:
Resolved Stellar Populations, Chavez M., Valls-Gabaud D., eds.,
(ASP: San Francisco), in press (astro-ph/0509798)
\bibitem[\protect\citeauthoryear{Chiosi} {2007}]{b4}
        Chiosi E., 2007, A\&A, 466, 165
\bibitem[\protect\citeauthoryear{Covey} {2008}]{b4} Covey K. R.,
        Hawley S. L., Bochanski J. J., West A. A., Reid I. N.,
        Golimowski D. A., Davenport J. R. A., Henry T., 2008, AJ, 136,
        1778
\bibitem[]{} Da Rio N., Gouliermis D. A., Henning T., 2009, ApJ, in
press (arXiv:0902.0758)
\bibitem[\protect\citeauthoryear{de Grijsa} {2002}]{b7}
        de Grijs R., Johnson R. A., Gilmore G. F., Frayn C. M., 2002a, MNRAS,
        331, 228
\bibitem[\protect\citeauthoryear{de Grijsb} {2002}]{b6}
        de Grijs R., Gilmore G. F., Johnson R. A., Mackey A. D., 2002b, MNRAS,
        331, 245
\bibitem[\protect\citeauthoryear{de Grijsc} {2002}]{b8} de Grijs R.,
        Gilmore G. F., Mackey A. D., Wilkinson M. I., Beaulieu S. F.,
        Johnson R. A., Santiago B. X., 2002c, MNRAS, 337, 597
\bibitem[\protect\citeauthoryear{Delfosse} {2004}]{b9}
        Delfosse X., Beuzit J.-L., Marchal L., Bonfils X., Perrier C.,
        S\'{e}gransan D., Udry S., Mayor M., Forveille T., 2004, in:
        Spectroscopically and Spatially Resolving the Components of
        Close Binary Stars, Hilditch R. W., Hensberge H., Pavlovski
        K., eds., ASP Conf. Ser., (ASP: Sab Francisco), Vol. 318,
        p. 166
\bibitem[\protect\citeauthoryear{Dotter et
al.}{2008}]{2008ApJS..178...89D} Dotter A., Chaboyer B.,
Jevremovi{\'c} D., Kostov V., Baron E., Ferguson J.~W., 2008, ApJS,
178, 89
\bibitem[\protect\citeauthoryear{Eisenhauer} {2001}]{b10} Eisenhauer
F., 2001, in: Starburst galaxies: near and far, Tacconi L., Lutz D.,
eds., (Springer: Heidelberg), p. 24
\bibitem[\protect\citeauthoryear{Elson} {1998}]{b9}
        Elson R., Sigurdsson S., Davies M., Hurley J., Gilmore G., 1998,
        MNRAS, 300, 857
\bibitem[\protect\citeauthoryear{Elson} {1999}]{b9} Elson R., Tanvir
        N., Gilmore G., Johnson R. A., Beaulieu S., 1999, in: New
        Views of the Magellanic Clouds, Chu Y.-H., Suntzeff N., Hesser
        J., Bohlender D., eds., IAU Symp. 190, (ASP: San Francisco),
        p. 417
\bibitem[\protect\citeauthoryear{Elson} {1989}]{b9}
        Elson R., Freeman K., Lauer T., 1989,
        ApJ, 347, 69
\bibitem[\protect\citeauthoryear{Gilmore} {2001}]{b10} Gilmore G.,
        2001, in: Starburst galaxies: near and far, Tacconi L., Lutz D.,
        eds., (Heidelberg: Springer), p. 34
\bibitem[\protect\citeauthoryear{Girardi} {2000}]{b10}
        Girardi L., Bressan A., Bertelli G., Chiosi C., 2000, A\&AS,
        141, 371
\bibitem[\protect\citeauthoryear{Gouliermis} {2006}]{b11}
        Gouliermis D., Brandner W., Henning Th., 2006a, ApJ, 636, L133
\bibitem[\protect\citeauthoryear{Gouliermis} {2006}]{b11}
        Gouliermis D., Brandner W., Henning Th., 2006b, ApJ, 641, 838
\bibitem[\protect\citeauthoryear{Hauschildt, Allard, \&
Baron}{1999}]{1999ApJ...512..377H} Hauschildt P.~H., Allard F., Baron
E., 1999, ApJ, 512, 377
\bibitem[\protect\citeauthoryear{Hennebelle \&
        Chabrier}{2008}]{2008ApJ...684..395H} Hennebelle P., Chabrier
        G., 2008, ApJ, 684, 395
\bibitem[\protect\citeauthoryear{Hillenbrand \&
White}{2004}]{2004ApJ...604..741H} Hillenbrand L.~A., White R.~J.,
2004, ApJ, 604, 741
\bibitem[\protect\citeauthoryear{Holtzman} {1995}]{b12} Holtzman
        J. A., Burrows C. J., Casertano S., Hester J. J., Trauger
        J. T., Watson A. M., Worthey G., 1995, PASP, 107, 1065
\bibitem[]{} Hu Y., Deng L., de Grijs R., Goodwin S. P., Liu Q., 2009,
        ApJ, submitted (arXiv:0801.2814)
\bibitem[]{} Hunter D. A., Light R. M., Holtzman J. A., Lynds R.,
        O'Neil E. J., Grillmair C. J., 1997, ApJ, 478, 124
\bibitem[\protect\citeauthoryear{Johnson} {2001}]{b13} Johnson R.A.,
        Beaulieu S. F., Gilmore G. F., Hurley J., Santiago B. X.,
        Tanvir N. R., Elson R. A. W., 2001, MNRAS, 324, 367
\bibitem[\protect\citeauthoryear{Kerber} {2006}]{b14}
        Kerber L. O., Santiago B. X., 2006, A\&A, 452,155
\bibitem[\protect\citeauthoryear{Krist}{2001}]{KH01} Krist, J., Hook,
        R., 2001, The Tiny Tim User's Guide, Version 6.0 ({\tt
        http://www.stsci.edu/software/tinytim/})
\bibitem[\protect\citeauthoryear{Kroupa} {2008}]{k08} Kroupa P., 2008,
in: The Cambridge $N$-body Lectures, Aarseth S., Tout C., Mardling R.,
eds., (Springer: Heidelberg), Lecture Notes in Physics, vol. 760,
p. 181
\bibitem[\protect\citeauthoryear{Kroupa} {1993}]{b14}
        Kroupa P., Tout C. A., Gilmore G. F., 1993, MNRAS, 262,545
\bibitem[\protect\citeauthoryear{Kroupa} {2001}]{b15}
        Kroupa P., 2001, MNRAS, 322, 231
\bibitem[\protect\citeauthoryear{Kroupa} {2003}]{b15}
        Kroupa P., Weidner C., 2003, ApJ, 598, 1076
\bibitem[\protect\citeauthoryear{Kroupa} {2007}]{b15} Kroupa P., 2007,
        in: Resolved Stellar Populations, Valls-Gabaud D., Chavez M.,
        eds., (ASP: San Francisco), in press (astro-ph/0703124)
\bibitem[\protect\citeauthoryear{Marchal} {2003}]{b9} Marchal
        L., Delfosse X., Forveille T., S\'{e}gransan D., Beuzit J.-L.,
        Udry S., Perrier C., Mayor M., Halbwachs J.-L., 2003, in:
        Brown Dwarfs, Mart\'{\i}n E., ed., IAU Symp. 211, p. 311
\bibitem[\protect\citeauthoryear{Mathieu} {2007}]{b15} Mathieu
        R. D., Baraffe I., Simon M., Stassun K. G., White R., 2007,
        Protostars and Planets V, Reipurth B., Jewitt D., Keil K.,
        eds., (Univ. of Arizona Press: Tucson), p. 411
\bibitem[\protect\citeauthoryear{Palla \&
Stahler}{1999}]{1999ApJ...525..772P} Palla F., Stahler S.~W., 1999,
ApJ, 525, 772
\bibitem[\protect\citeauthoryear{Paresce} {2000}]{b16}
        Paresce F., De Marchi G., 2000, ApJ, 534, 870
\bibitem[\protect\citeauthoryear{Park} {2000}]{b17}
        Park B., Sung H., Bessell M., Kang Y., 2000, AJ, 120, 894
\bibitem[\protect\citeauthoryear{Salpeter} {1955}]{b19}
        Salpeter E. E., 1955, AJ, 101, 1865
\bibitem[\protect\citeauthoryear{Santiago} {2001}]{b20}
        Santiago B. X., Beaulieu S., Johnson R., Gilmore G. F., 2001, A\&A,
        369, 74
\bibitem[\protect\citeauthoryear{Scalo} {1998}]{b9} Scalo J., 1998,
in: The stellar initial mass function, Gilmore G., Howell D., eds.,
ASP Conf. Ser., (ASP: San Francisco), vol. 142, p. 201
\bibitem[scalo05]{2005} Scalo J., 2005, in: The initial mass function
50 years later, Corbelli E., Palle F., Zinnecker H., eds., ApSS
Library, vol. 327, p. 23
\bibitem[\protect\citeauthoryear{Siess} {2000}]{b20}
        Siess L., Dufour E., Forestini M., 2000, A\&A, 358, 593
\bibitem[\protect\citeauthoryear{Sung} {2000}]{b18}
        Sung H., Chun M., Bessell M., 2000, AJ, 120, 333
\bibitem[\protect\citeauthoryear{Tout} {96}]{b21}
        Tout C. A., Pols O. R., Eggleton P. P., Han Z., 1996, MNRAS,
        281, 257
\bibitem[\protect\citeauthoryear{van den
        Bergh}{1981}]{1981A&AS...46...79V} van den Bergh S., 1981,
        A\&AS, 46, 79
\bibitem[\protect\citeauthoryear{White} {1999}]{b22}
        White R. J., Ghez A. M., Reid I. N., Schultz G., 1999,
        ApJ, 520, 811
\bibitem[\protect\citeauthoryear{Whitmore} {1999}]{b23}
        Whitmore B., Heyer I., Casertano S., 1999, PASP,
        111, 1559
\bibitem[\protect\citeauthoryear{Will} {1995}]{b24}
        Will J. M., Bomans D.J., de Boer K.S., 1995, A\&A, 295, 54
\bibitem[\protect\citeauthoryear{Yi, Kim, \&
Demarque}{2003}]{2003ApJS..144..259Y} Yi S.~K., Kim Y.-C., Demarque
P., 2003, ApJS, 144, 259

\end{thebibliography}
\end{document}